\documentclass[10pt]{article} 
\usepackage[accepted]{rlc}
\usepackage{booktabs}
\usepackage{float}
\usepackage{amssymb}            
\usepackage{mathtools}          
\usepackage{mathrsfs}           
\mathtoolsset{showonlyrefs}     
\usepackage{graphicx}           
\usepackage{subfigure}
\graphicspath{{./figures/}}
\usepackage{subcaption}         
\usepackage[space]{grffile}     
\usepackage{url}                

\title{
    \centering
    \begin{minipage}{0.85\textwidth} 
        \centering
        \textbf{SoDA: An Efficient Interaction Paradigm for the Agentic Web}
    \end{minipage}
}

\author{
    \hspace*{-2em}
    \begin{minipage}{\linewidth}
        \centering
        \textbf{
            Zicai Cui\textsuperscript{1}, 
            Yuanjian Zhou\textsuperscript{2}, 
            Weiwen Liu\textsuperscript{1}, 
            Weinan Zhang\textsuperscript{1,2*} 
        }
        \\[1em] 
        \normalfont
        \textsuperscript{1}Shanghai Jiao Tong University \quad
        \textsuperscript{2}Shanghai Innovation Institute 
        \\[0.5em] 
        \texttt{zicaicui@sjtu.edu.cn} \quad 
        \texttt{jake.zhou@sii.edu.cn} \\
        \texttt{wwliu@sjtu.edu.cn} \quad 
        \texttt{wnzhang@sjtu.edu.cn}
    \end{minipage}
}

\begin{document}

\maketitle

\begin{abstract}
As the internet evolves from the mobile App-dominated Attention Economy to the Intent-Interconnection of the \textit{Agentic Web} era, existing interaction modes fail to address the escalating challenges of data lock-in and cognitive overload. Addressing this challenge, this paper defines a future-oriented user sovereignty interaction paradigm, aiming to realize a fundamental shift from killing time to saving time. Specifically, we argue that decoupling memory from application logic eliminates the structural basis of data lock-in, while shifting from explicit manual instruction to implicit intent alignment resolves cognitive overload by offloading execution complexity. This paradigm is materialized through the construction of the \textbf{Sovereign Digital Avatar (SoDA)}, which employs an orthogonal decoupling design of storage, computation, and interaction. This establishes the architectural principle of data as a persistent asset, model as a transient tool, fundamentally breaking the platform monopoly on user memory. To support the operation of this new paradigm in zero-trust environments, we design an \textit{Intent-Permission Handshake Mechanism} based on A2A protocols, utilizing dual-factor (Sensitivity Coefficient and Strictness Parameter) adaptive routing to achieve active risk governance. Empirical evaluation with a high-fidelity simulation environment indicates that this paradigm reduces token consumption by approximately 27-35\% during cross-platform service migration and complex task execution. Furthermore, in the orchestration of multi-modal complex tasks, it reduces user cognitive load by 72\% compared to standard standard Retrieval-Augmented Generation (RAG) architectures, by 88\% relative to manual workflows, while significantly boosting the Information Signal-to-Noise Ratio (SNR). These results demonstrate that the SoDA is the essential interaction infrastructure for building an efficient, low-friction, and decentralized Agentic Web.
\end{abstract}

\section{Introduction}
\label{sec:1}

The evolution from Mobile Web to the Agentic Web \citep{yang2025agentic} signifies a fundamental paradigm shift in how humans interact with the digital world. In the Mobile Web era, user interaction is largely fragmented, structured around isolated, app-based experiences. To capture user attention, platforms have constructed closed walled gardens, utilizing algorithmic recommendation mechanisms to lock users into immersive content consumption oriented towards \textit{Killing Time} \citep{wu2017attention, zuboff2019age}. While this model has achieved immense commercial success, it has resulted in extreme information fragmentation and the excessive consumption of user cognitive resources \citep{ellingrud2023generative}.

With the explosive growth of Large Language Model (LLM) capabilities, the internet is entering the Agentic Web era \citep{yang2025agentic,xi2025rise}. As noted in recent research  \citep{wei2025ai}, LLMs offer unprecedented opportunities to redefine memory management and interaction methods, making the construction of AI-native personal memory systems possible. Under this new paradigm, the subject of interaction shifts from Human-App to the collaboration of Human-Agent and Agent-Agent. Future interaction will no longer require users to frequently switch between different applications; instead, knowledge exchange with the external world will occur through personal intelligent agents. The core vision of this transformation is a shift from passive attention consumption to active intent execution. Crucially, this paradigm does not seek to negate immersive entertainment, but to reclaim the time squandered on functional friction. It introduces a \textit{Saving Time} alternative for goal-oriented tasks, ensuring that users are no longer forced to endure \textit{Killing Time} mechanisms just to migrate data or complete forms. Instead of navigating complex interfaces designed to capture attention, users can achieve goals directly through agents, liberating humans from redundant digital interaction, liberating humans from tedious digital interactions through intelligent task automation and information filtering. 

\begin{figure}[!h]
  \centering
  \includegraphics[width=0.8\textwidth]{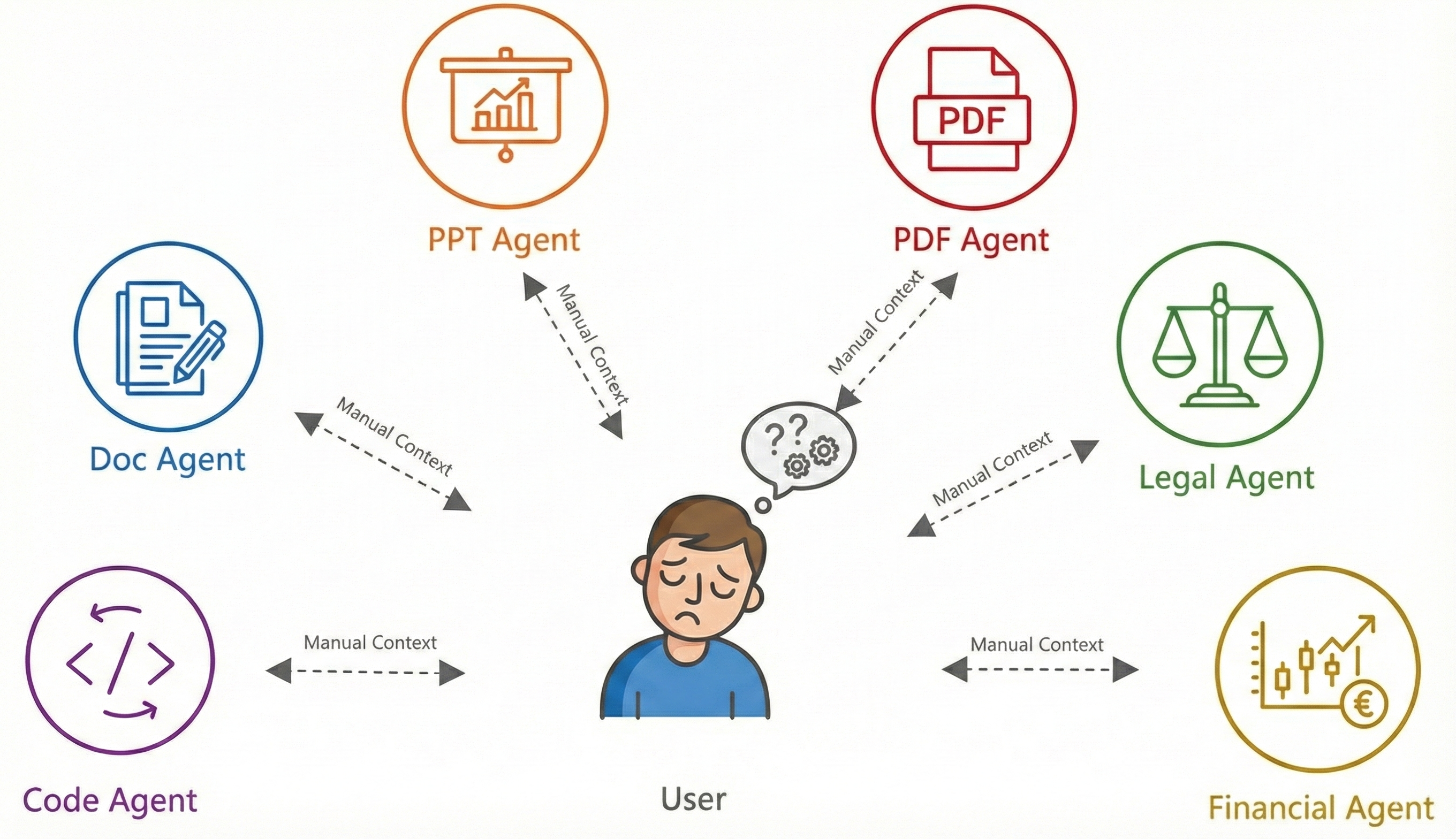}
  \caption{The Human-as-Router Dilemma. Users face high cognitive load when manually coordinating context across isolated vertical agents, leading to severe information fragmentation.}
  \label{figure1}
\end{figure}

Despite the promising future depicted by the Agentic Web, current implementation architectures still face two core challenges:

\begin{itemize}
    \item \textbf{Data Lock-in:} Mainstream AI agents (e.g., ChatGPT, Claude) often strongly bind user data to specific model parameters or platform ecosystems. Existing solutions, such as browser-stored credentials or unified authentication systems, mitigate redundancy to some extent but typically exist as static storage, lacking contextual reasoning capabilities and adaptability. More critically, users do not truly possess a platform-independent portable memory bank. Recent reviews on Agentic AI highlight that privacy and data ownership remain the primary bottlenecks hindering widespread adoption in  \cite{bandi2025rise}. Once a service provider is changed, long-accumulated personalized data and interaction habits face the risk of loss  \citep{mansour2016demonstration}.
    \item \textbf{Cognitive Overload:} In the nascent stage of multi-agent coexistence, users are forced to act as routers, manually transferring information between agents in different vertical domains (e.g., search, image generation, and writing Agents). This frequent context switching and repetitive information entry lead to severe cognitive fatigue, hindering the seamless integration of technology and users. Although existing research has attempted to use systems like Second Me  \citep{wei2025ai} as an intermediary layer to reduce interaction friction, achieving complete data decoupling at the architectural level and establishing a quantifiable standard to evaluate interaction efficiency improvements remain urgent problems to be solved.
\end{itemize}

To fundamentally dismantle data lock-in, our architecture implements an orthogonal decoupling of storage and computation. By encapsulating user history and preferences into a portable SMP, we transform memory from a platform-bound dependency into a user-owned persistent asset \citep{packer2023memgpt}. This ensures that the digital avatar can migrate seamlessly across diverse services without context loss. Simultaneously, to alleviate \textit{Cognitive Overload} \citep{risko2016cognitive}, we shift the interaction paradigm from explicit manual orchestration to implicit intent alignment. The avatar acts as an intelligent proxy, utilizing an \textit{Intent-Permission Handshake} \citep{shneiderman2022human} to autonomously filter noise and negotiate permissions, thereby achieving cognitive offloading that liberates users from low-level execution details.

Addressing the aforementioned issues, this paper proposes a \textit{Sovereignty Architecture} for the Agentic Web. The core vehicle of this architecture is the SoDA, which is an intelligent agent constructed on data decoupling technology that fully represents user intent.The main contributions of this paper are as follows:

\begin{itemize}
    \item \textbf{Proposal of a Fully Decoupled Data Architecture:} Unlike pathways that internalize memory into model parameters, we design a mechanism separating storage from computation. This ensures that the user's digital memory can be independently extracted, migrated, and losslessly reused in heterogeneous Agent environments, realizing true data sovereignty.
    \item \textbf{Construction of an Intent-based Interaction Shielding Mechanism:} We design the Digital Avatar as an intelligent shield between the user and the external Agentic Web. Utilizing a local memory bank, it autonomously handles redundant interactions, triggering Human-in-the-loop (HITL)  \citep{wu2022survey} confirmation only at critical nodes, thereby maximizing the reduction of user cognitive load.
    \item \textbf{Quantitative Evaluation based on HCI Models:} Distinct from evaluation systems relying solely on LLM scoring, this paper introduces the Keystroke-Level Model(KLM) \citep{card1980keystroke} from the Human-Computer Interaction(HCI)\citep{preece1994human} domain and the SNR from information theory. Through simulation experiments, we quantitatively demonstrate the significant advantages of this new paradigm in terms of interaction efficiency and cognitive offloading.
\end{itemize}

\section{Related Work}
\label{sec:2}

\subsection{Agentic Web and Personal Agents}
With the rapid development of LLMs, personal AI agents have become critical components in achieving Artificial General Intelligence (AGI) \citep{goertzel2007artificial}. Early solutions focused primarily on rule-based automation (e.g., Auto-fill), lacking the ability to understand context.

Recent research has begun to explore AI-native memory paradigms  \citep{zhong2024memorybank}. For instance, \cite{wei2025ai} propose a hybrid architecture comprising L0 (Raw Data), L1 (Natural Language Memory), and L2 (AI-Native Memory) layers. By parameterizing memory, the system enables the model to act as an extension of the user, automatically generating context-aware responses and pre-filling information. Such work demonstrates the immense potential of personal agents as intermediaries between users and digital ecosystems  \citep{park2023generative}. However, most existing work (including LPM 1.0) focuses more on enhancing the model's internalization of memory through Supervised Fine-Tuning (SFT) and Direct Preference Optimization (DPO) \citep{rafailov2023direct}. In contrast, this paper focuses on architectural data independence and cross-model universality.

\subsection{Data Sovereignty and Decoupling}
Data sovereignty refers to users having complete control over their digital assets. In traditional Retrieval-Augmented Generation (RAG) or RALM architectures, although data exists in unstructured forms, it often relies on specific retrievers and vector databases, making migration between platforms difficult  \citep{ram2023context}.

Related initiatives, such as the Solid project proposed by Tim Berners-Lee, advocate storing data in independent pods  \citep{sambra2016solid}, but this concept has not yet been fully integrated and validated within the context of the AI era. Unlike \cite{wei2025ai} which attempts to organize and learn L2 layer memory through neural network parameters, this paper advocates for a more explicit data decoupling solution based on standardized protocols. We argue that in the Agentic Web, data should not merely be memorized by a model but should be capable of being carried by the user and mounted onto any authorized computing node, thereby avoiding the lock-in risks associated with catastrophic forgetting  \citep{luo2025empirical}.

\subsection{HCI Evaluation Models}
Evaluating the efficacy of personal intelligent agents is a complex issue. Current evaluation methods predominantly rely on automated metrics or LLM scoring  \citep{liu2023g}. For example, \cite{wei2025ai} study introduced an LLM-based automated evaluation pipeline covering memory QA, context enhancement, and context critic tasks, utilizing metrics such as helpfulness, completeness, and empathy for scoring. While these content-based evaluations reflect answer quality, they often overlook the cost of the interaction process itself.

As noted in related research, LLM evaluations may underestimate the quality of actual experiences and often exhibit a bias towards longer texts  \citep{zheng2023judging}. To bridge this gap, this paper introduces classic models from the HCI domain, such as GOMS (Goals, Operators, Methods, and Selection rules) and the Keystroke-Level Model (KLM)  \citep{card2018psychology}. By calculating context switching counts and operational time costs, we aim to quantify the actual contribution of the Agentic Web paradigm to reducing user interaction friction from the perspective of \textit{cognitive Economics}, supplementing existing content-based evaluation systems.

\section{System Architecture: A User Sovereignty Model for the Agentic Web}
\label{sec:3}

The section propose the \textbf{user sovereignty model}, a decentralized framework designed to secure user context within the Agentic Web. The architecture is founded on a tripartite design that orthogonally decouples \textit{Storage}, \textit{Computation}, and \textit{Interaction}, treating data as a persistent asset independent of the computational model. This section presents the system's comprehensive design, moving from the macroscopic topological structure to the microscopic SMP, and culminating in the \textit{Intent-Permission Handshake} mechanism for active risk governance.

\subsection{Orthogonally Decoupled Three-Layer Architecture}

To address the structural challenges of data lock-in and cognitive overload within the current Agent ecosystem, this paper proposes a \textbf{SoDA Architecture} based on the principle of \textit{Orthogonal Separation of Concerns}. Our architecture establishes a design paradigm of data as persistent asset, model as transient tool. By orthogonal, we imply that the lifecycles of these core components are functionally independent: the persistence of user memory is no longer contingent upon the specificity of a computational model, nor is it bound by the proprietary logic of an interaction platform. This structural independence ensures that upgrading the reasoning engine or migrating across services does not necessitate the reconstruction of the user's digital context. Based on this, we construct a three-layer decoupled model comprising storage, computation, and interaction:

\begin{figure}[H]
  \centering
  \includegraphics[width=0.9\textwidth]{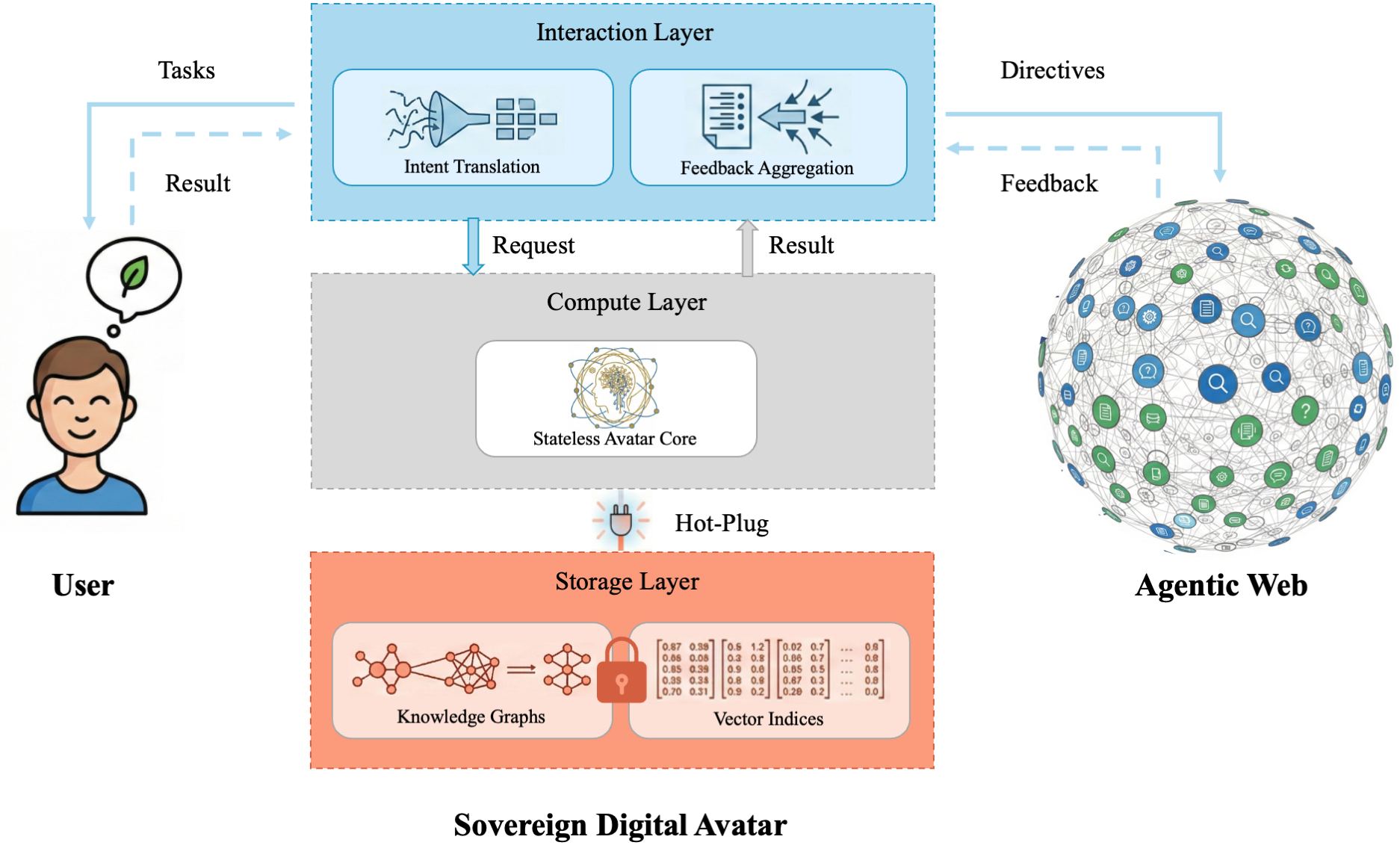}
  \caption{The Orthogonally Decoupled Architecture of the SoDA. The system is structured into three functionally independent layers: (Top) Interaction Layer, which translates ambiguous user Intents into structured Directives for the Agentic Web and aggregates external Feedback into a low-cognitive-load Briefing; (Middle) Compute Layer, a stateless runtime that performs reasoning without data retention; and (Bottom) Storage Layer, the Sovereign Memory Pod (SMP) housing hybrid Knowledge Graph and Vector assets. A key innovation is the Hot-Plug mechanism, allowing the model to transiently access persistent data, ensuring user sovereignty by preventing data lock-in.}
  \label{figure2}
\end{figure}

The architecture is composed of three distinct, loosely coupled layers(Figure \ref{figure2}), ensuring that the user's digital existence is not tethered to any single model provider or platform:

\textbf{Storage Layer:} Sovereign Memory Pod (SMP). This layer constitutes the physical carrier of user digital sovereignty, formalized as an encrypted, heterogeneous data container. It does not rely on specific neural network architectures but employs a hybrid storage mechanism designed for semantic completeness. Knowledge Graphs \citep{hogan2021knowledge} are utilized to explicitly store determinative user attributes (e.g., social connections, asset portfolios) and their topological relationships, enabling multi-hop reasoning for complex intent deduction; simultaneously, vector indices are employed to encode high-dimensional embeddings of unstructured interaction logs (e.g., chat history, browsing footprints), capturing fuzzy semantic nuances. This hybrid topology leverages the structural precision of graphs and the semantic flexibility of vectors, ensuring that the system can retrieve both exact facts and latent contexts—a capability consistent with the \textit{GraphRAG} paradigm that demonstrates significantly higher recall accuracy for complex queries compared to naive vector retrieval \citep{edge2024local}. The SMP features independent addressability and portability, ensuring data exists physically independent of any inference model.

\textbf{Compute Layer:} Stateless Avatar Core. This layer functions as the execution runtime of the user's \textit{Digital Avatar}. Acting as the user's digital twin, it is the exclusive entity authorized with the permissions to access and retrieve knowledge from the encrypted SMP. Unlike traditional models that lock memory within their parameters, the \textit{Avatar Core} operates as a stateless reasoning agent. It dynamically accesses the SMP to retrieve relevant context, utilizing this private knowledge to bridge the gap between user intent and external execution. This mechanism enables the avatar to effectively orchestrate complex interactions with external agents in the Agentic Web, ensuring that while the model leverages user data for reasoning, it never retains ownership of it.

\textbf{Interaction Layer:} Intent-Proxy Interface. This layer serves as the standardized gateway to the external Agentic Web. It encapsulates the complexity of internal memory retrieval and reasoning, exposing a unified Agent-to-Agent (A2A) communication interface to the outside world  \citep{ray2025review}. Its core function is to translate the user's unstructured, ambiguous intents into structured Directives executable by external agents, and to aggregate heterogeneous external feedback into user briefings with low cognitive load.

\subsection{A2A-based Intent-Permission Handshake and Adaptive Routing}

\begin{figure}[H]
  \centering
  \includegraphics[width=0.9\textwidth]{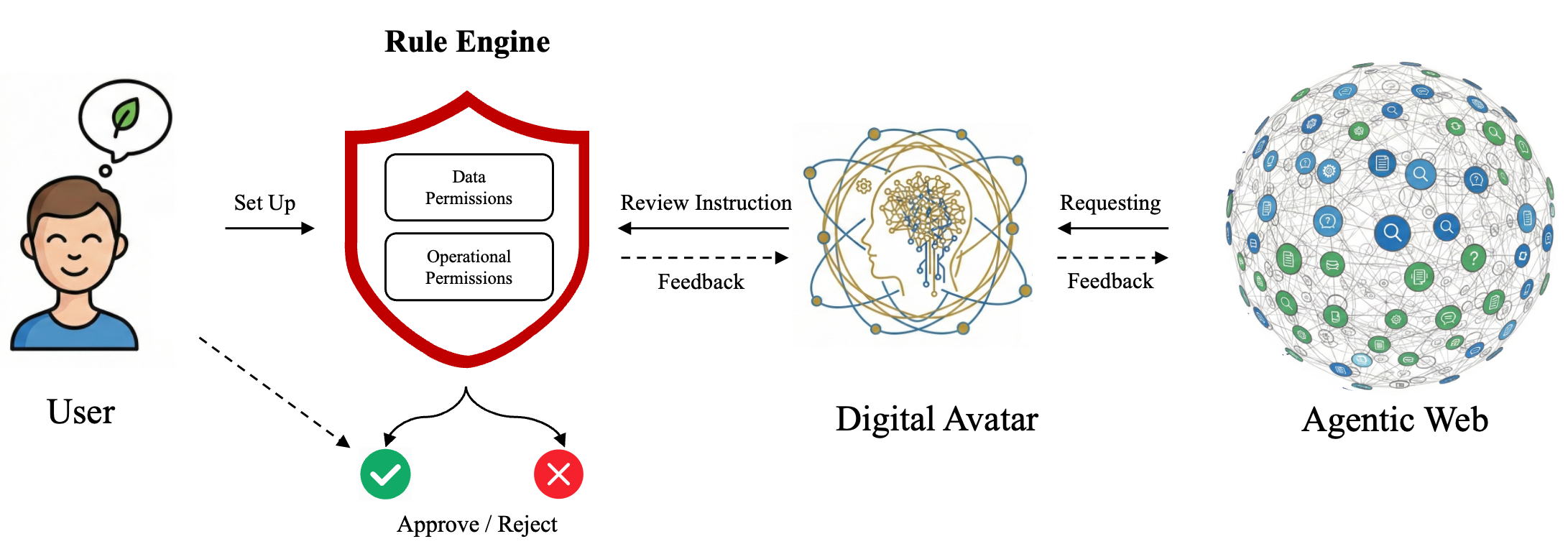}
  \caption{The Intent-Permission Handshake Mechanism. A rule engine acts as a privacy shield, dynamically approving or rejecting external agent requests based on user-defined policies and operational permissions.}
  \label{fig:handshake}
\end{figure}

In the open and Zero-Trust environment of the Agentic Web  \citep{owasp2023owasp}, the primary responsibility of the \textit{Digital Avatar} is to establish privacy boundaries while ensuring interaction efficiency. To this end, we propose an \textit{Intent-Permission Handshake Protocol} built upon standard agent communication protocols.  As shown in Figure \ref{fig:handshake}, the \textit{Intent-Permission Handshake Mechanism} functions as an active privacy firewall. The rule engine dynamically evaluates external requests against user-defined data and operational permissions, ensuring that only auth.

When an external agent initiates an interaction request, the Avatar first issues a semantic challenge, verifying the minimum necessity of the counterparty's intent via an embedded parser. Subsequently, the system executes a quantitative assessment, mapping the requested fields to the Universal Profile Description Language(UPDL) ontology layer within the SMP to calculate an objective \textit{Request Sensitivity Coefficient ($R$)}. Based on this, the system introduces a dual-factor decision mechanism, combining the user-defined subjective \textit{Strictness Parameter ($S$)} to construct a joint routing function $\Phi(S, R)$. This function dynamically maps interaction requests to one of three response spaces:

\begin{itemize}
    \item \textbf{Auto-Zone:} When $\Phi(S, R)$ is below a preset safety threshold (i.e., the request has low sensitivity and aligns with user preset policies), the system autonomously completes data retrieval and response with zero cognitive load on the user. This mechanism constitutes the core pathway for realizing the save time paradigm, effectively filtering out massive amounts of basic handshake noise.
    \item \textbf{Negotiation-Zone:} When a request involves high-value privacy assets or reaches a critical threshold, the avatar initiates a defensive Negotiation Subroutine. The system automatically performs differential privacy processing, such as data granularity scaling, or requires the external agent to provide a Proof of Value. If a consensus cannot be reached, the system escalates to trigger a HITL binary confirmation mechanism  \citep{mosqueira2023human}.
    \item \textbf{Blocking-Zone:} For unauthorized intents, anomalous traffic patterns, or requests touching the highest sensitivity threshold, the system executes an immediate circuit break, returning a standardized rejection signal and logging the event for audit. This mathematically constructs a digital firewall based on Zero-Trust principles.
\end{itemize}

\subsection{Data Hot-Plugging and Semantic Interoperability}

To realize the vision of data truly following the user rather than being tethered to a specific platform, this architecture implements a decoupling mechanism based on \textit{Data Hot-Plugging}. Unlike static knowledge bases, the SMP is designed as a portable, standardized artifact. At runtime, the \textit{Avatar Core} establishes a transient secure connection with the SMP via RAG technology, loading high-frequency context into working memory  \citep{gao2024modular}. Upon session termination or service migration, the system executes a mandatory unmount operation, thoroughly erasing the local context window. This Burn-after-reading characteristic guarantees forward secrecy between sessions, fundamentally eliminating the risk of vendor lock-in  \citep{li2023privacy}.

Furthermore, to ensure the readability of decoupled data in heterogeneous environments, we define the Universal Profile Description Language(UPDL). Formally, UPDL is a schema-agnostic serialization protocol based on the JSON-LD standard, designed to map discrete user attributes into a unified Knowledge Graph. By leveraging standardized ontologies to build a semantic network, UPDL encapsulates user preferences, historical interactions, and behavioral patterns as self-describing, platform-independent semantic nodes. This standardized \textit{Intermediate Representation} allows the SMP to flow seamlessly between different Agentic systems—when a user migrates from one service domain to another, the target agent can instantly construct a User context model by parsing the UPDL graph without cold-start training or parameter updates. This mechanism not only safeguards the physical independence of user data but also achieves semantic interoperability across ecosystems  \citep{balog2019personal}, providing a solid technical implementation path for data sovereignty in the Agentic Web era.

\section{Experiments and Evaluation}
To validate the effectiveness of the \textit{SoDA Architecture} in enhancing interaction efficiency, guaranteeing data sovereignty, and reducing cognitive load, we constructed a high-fidelity simulation environment containing heterogeneous service agents and designed three progressive controlled experiments. Unlike existing studies that predominantly rely on LLM subjective scoring (LLM-as-a-judge) \citep{li2025generation}, this study introduces classic cognitive models from HCI and information-theoretic metrics. Our goal is to mathematically quantify the performance gains of this new Agentic Web paradigm relative to traditional interaction modes. Three research questions (RQ) lead the following discussions.
\begin{itemize}
    \item \textbf{RQ1 (Interaction Efficiency):} Can the SoDA architecture significantly reduce user cognitive load and interaction friction in complex, multi-agent task orchestration compared to traditional paradigms?
    \item \textbf{RQ2 (Data Sovereignty):} How does the orthogonally decoupled storage-computation architecture impact the efficiency of cross-platform service migration and the cold start problem?
    \item \textbf{RQ3 (Privacy Governance):} Can the intent-permission handshake mechanism effectively balance the trade-off between privacy protection and service utility in a Zero-Trust environment?
\end{itemize}

\subsection{Experimental Setup}

To reproduce the complexity of the Agentic Web while ensuring verifiable metrics, we developed a hybrid simulation environment. This framework integrates a Python-based discrete-event engine for orchestrating asynchronous interaction flows (e.g., network latency, concurrency) with real-time LLM inference endpoints (GPT-4o).

\paragraph{Task Benchmarks.} 
To rigorously evaluate the agent's capability across distinct dimensions of user sovereignty, we designed a benchmark suite consisting of \textbf{four representative tasks}. These tasks categorize into three types:
\begin{itemize}
    \item \textbf{Type I: Context-Aware Information Processing} (e.g., Task 1: Smart Video-to-Note, Task 2: Intelligent Literature Screening). These tasks test the agent's ability to filter massive external information based on internal user preferences.
    \item \textbf{Type II: Risk-Sensitive Governance} (e.g., Task 3: Local Code Security Audit). This task specifically evaluates the agent's autonomous behavior in identifying privacy leaks (e.g., API keys) and enforcing security policies in a zero-trust environment.
    \item \textbf{Type III: Latent Intent Inference} (e.g., Task 4: Personalized Financial Dashboard). These tasks require the agent to infer implicit user goals (e.g., investment interests) from heterogeneous profile data rather than explicit instructions.
\end{itemize}

Within this environment, we utilize a standardized user simulator loaded with a predefined SMP to ensure consistency. We compare four distinct interaction paradigms:

\begin{itemize}
    \item \textbf{Manual:} The user manually operates tools or interfaces (Human-as-Router), representing the upper bound of accuracy but the lower bound of efficiency.
    \item \textbf{General Agent:} The user interacts with a standard ReAct-based agent without external memory access, representing the factory setting of current LLMs.
    \item \textbf{Strong RAG Agent:} An advanced agent equipped with the same toolset as Ours but utilizing a standard vector database for retrieval-augmented generation. This represents the current industry state-of-the-art for memory-augmented agents.
    \item \textbf{SoDA(Ours):} The user interacts through the \textbf{SoDA} equipped with the SMP architecture, capable of full context awareness and autonomous permission governance.
\end{itemize}

\subsection{Evaluation Metrics}
To comprehensively assess interaction efficiency and quality, we employ a set of multi-dimensional metrics. Specifically, we introduce quantitative models from HCI and Information Theory to strictly define Cognitive Load and Information Density.

\begin{itemize}
    \item \textbf{Cognitive Load Score ($L_{cog}$):} Based on the GOMS/KLM proposed by Card et al., we decompose the interaction process into primitive operations: Mental Preparation ($T_M$), Keystroking ($T_K$), Pointing ($T_P$), Homing ($T_H$), and System Waiting ($T_W$). The total cognitive load is quantified as the sum of time costs for all atomic operations:
    \begin{equation}
        L_{cog} = \sum_{i=1}^{N} (T_{M_i} + T_{K_i} + T_{P_i} + T_{H_i} + T_{W_i})
    \end{equation}
    
    \item \textbf{Interaction Friction Coefficient ($\eta$)}:
    A normalized metric [0, 1] quantifying the intensity of manual intervention required. It is defined as a weighted normalization of interaction turns ($N_{turns}$), mouse clicks ($N_{clicks}$), and text inputs ($N_{inputs}$):
    \begin{equation}
        \eta = \text{Normalize}(\alpha \cdot N_{turns} + \beta \cdot N_{clicks} + \gamma \cdot N_{inputs})
    \end{equation}
    where $\eta \rightarrow 0$ implies a fully autonomous \textit{Zero-Touch} workflow, while $\eta \rightarrow 1$ represents a completely manual operation.

    \item \textbf{Information Signal-to-Noise Ratio (SNR):} We define SNR to measure the effective information density delivered to the user. It is calculated as the ratio of tokens that strictly fulfill the user's intent to the total tokens the user is exposed to:
    \begin{equation}
        SNR = \frac{Tokens_{useful}}{Tokens_{total\_exposed}}
    \end{equation}
    
    \item \textbf{Task Completion Rate}: The percentage of tasks where the agent successfully fulfills the user's intent without critical errors or safety violations.
    
    \item \textbf{Result Deviation}: Measures the semantic discrepancy between the agent's output and the user's optimal expectation, specifically evaluating the accuracy of personalization.
    
    \item \textbf{Semantic Consistency \& Factual Recall}: Evaluates the agent's ability to maintain long-term context and accurately retrieve factual data (e.g., identity, assets) from the SMP across sessions.
    
    \item \textbf{Token Consumption}: The total number of tokens processed (input + output) to achieve the final result, serving as a direct proxy for economic cost and computational efficiency.
    
\end{itemize}
    
\subsection{Comprehensive Evaluation across Multi-Modal and Cross-Domain Scenarios (RQ1)}
To comprehensively validate the generalizability and orchestration efficacy of the \textit{SoDA} architecture in heterogeneous tasks, this study selected four representative task scenarios for comprehensive evaluation: (1) Task 1: Video-to-Note: Converting learning videos into academic notes; (2) Task 2: Paper Filtering (Domain Knowledge): Screening literature from a database; (3) Task 3: Code Audit (Privacy/Security): Detecting key leaks and verifying ownership; and (4)Task 4: FinTech Dashboard (Implicit Reasoning): Constructing a financial monitor based on user assets.

\begin{table}[H]
\centering
\footnotesize
\caption{Performance Comparison on Complex Task Orchestration}
\label{tab:complex_tasks}
\begin{tabular}{@{}lllll@{}}
\toprule
\textbf{Core Metric} & \textbf{Manual} & \textbf{General Agent} & \textbf{Strong RAG Agent} & \textbf{Digital Avatar(Ours)} \\
& \textit{(Manual)} & \textit{(General Agent)} & \textit{(Strong RAG Agent)} & \textit{(Digital Avatar(Ours)} \\ \midrule
Completion Rate & 100\% & 62.5\% & 81.25\% & \textbf{93.75\%} \\
Interaction Friction ($\eta$) & 0.925 & 0.55 & 0.35 & \textbf{0.05} \\
Cognitive Load Score ($L_{cog}$) & 8.6 & 4.6 & 3.8 & \textbf{1.05} \\
Information SNR & 14.05 & 5.0 & 18.5 & \textbf{29.9} \\
Total Token Consumption & N/A & 4,616 & 4120 & \textbf{2,989} \\ \bottomrule
\end{tabular}
\end{table}

\begin{figure}[H]
    \centering
    
    \noindent
    \begin{minipage}{\linewidth}
        \centering
        \includegraphics[width=\linewidth, height=0.23\textheight, keepaspectratio=true]{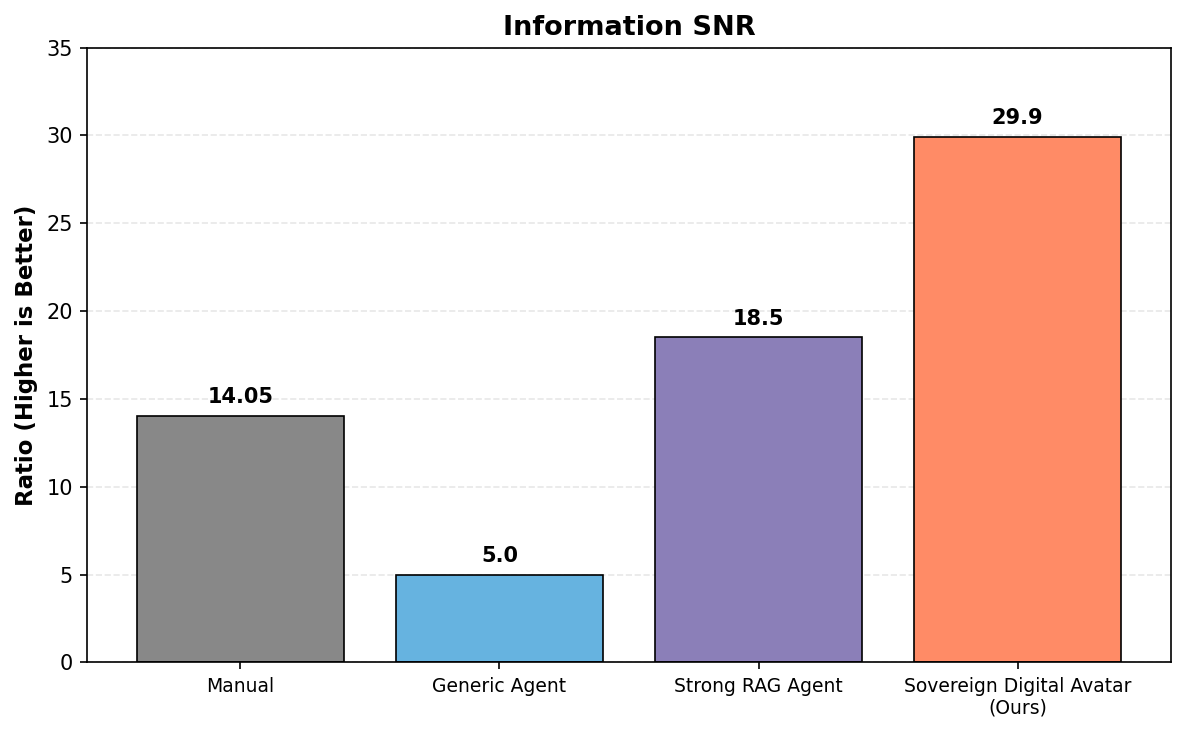} 
        \centerline{\small (a) The result of Information SNR} 
    \end{minipage}
    
    \vspace{0.5em} 
    
    \noindent
    \begin{minipage}{\linewidth}
        \centering
        \includegraphics[width=\linewidth, height=0.23\textheight, keepaspectratio=true]{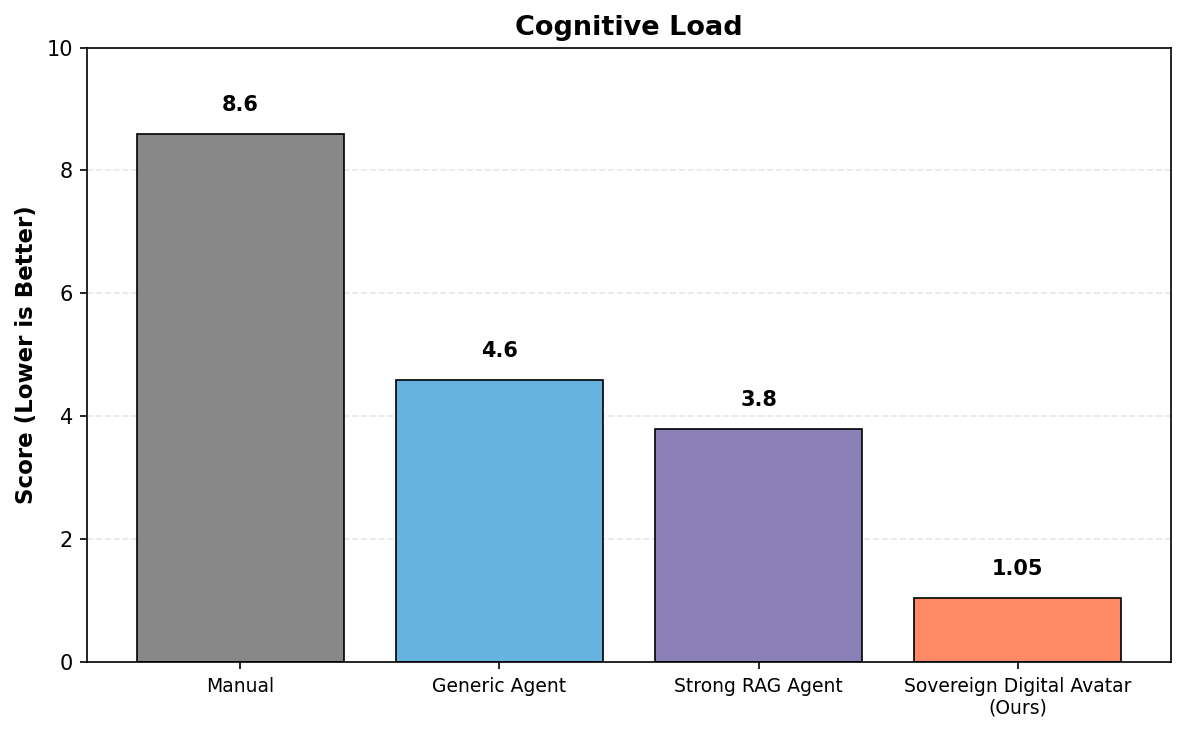}
        \centerline{\small (b) The result of Cognitive Load}
    \end{minipage}
    
    \vspace{0.5em} 
    
    \noindent
    \begin{minipage}{\linewidth}
        \centering
        \includegraphics[width=\linewidth, height=0.23\textheight, keepaspectratio=true]{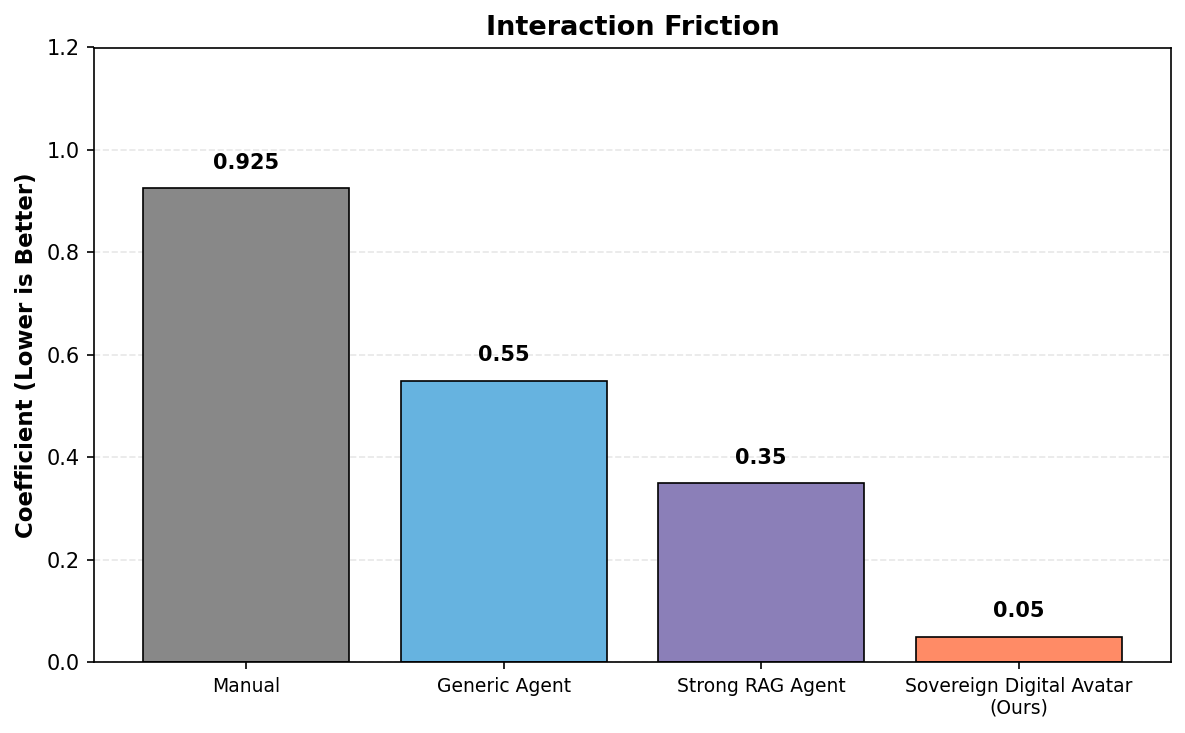}
        \centerline{\small (c) The result of Interaction Friction}
    \end{minipage}

    \caption{Quantitative Evaluation of Interaction Efficiency. Comparative analysis of SNR, Cognitive Load ($L_{cog}$), and Interaction Friction ($\eta$) across three paradigms. The SoDA (Ours) demonstrates significant performance gains over Manual and General Agent.}
    \label{fig:exp3_results}
\end{figure}

Table \ref{tab:complex_tasks} and Figure \ref{fig:exp3_results} present a comprehensive performance comparison. While the Strong RAG Agent achieves a completion rate of 81.25\%, it incurs high computational overhead. In contrast, our SMP-driven SoDA outperforms Strong RAG Agent in intent alignment while simultaneously optimizing efficiency—reducing friction by 85.7\% and token consumption by 27.5\%. Furthermore, the Digital Avatar demonstrates significant cognitive advantages over state-of-the-art baselines. Specifically, it reduces the Cognitive Load Score by 72.4\% compared to Strong RAG Agent, and by 77\% compared to General Agent, providing empirical validation of the system's capability."

\subsection{Evaluation of Data Decoupling and Cold Start Efficiency (RQ2)}

In the current Agentic Web ecosystem, user memory is often platform-locked. Moving from one platform to another necessitates a costly cold start. This experiment validates the portable memory mechanism of our SoDA.

\textbf{Scenario:} A user migrates to a new Agent platform and requests the generation of a personalized self-introduction. The content must strictly include: true name, status, research focus, representative papers, and hobbies.

\textbf{Results:} We compared the manual injection of context versus the structured injection of SMP (Ours). As shown in Table \ref{tab:coldstart}, the proposed architecture demonstrates significant efficiency gains.

\begin{table}[H]
\centering
\caption{Comparison of Cross-Platform Migration Efficiency }
\label{tab:coldstart}
\begin{tabular}{@{}lll@{}}
\toprule
\textbf{Metric} & \textbf{Manual} & \textbf{Memory Injection(Ours)} \\ \midrule
Cumulative Token Cost & 3,463 & \textbf{2,363} \\
Interaction Turns & 4 & \textbf{1} \\
Input Characters (Context) & 282 (Typing) & \textbf{0} (One-click) \\
Time to Result & 25.21s & \textbf{11.38s} \\ \bottomrule
\end{tabular}
\end{table}

The Baseline required 4 rounds of dialogue to correct hallucinations and supplement missing details, resulting in 3463 tokens consumed. In contrast, Ours achieved the goal in a single turn with zero character input from the user, effectively reducing the interaction friction to zero.

\subsection{Robustness Analysis of the Dual-Factor Privacy Gatekeeper (RQ3)}

In the Zero-Trust Agentic Web, ensuring privacy without sacrificing service utility is critical. We designed this experiment to validate the \textit{Intent-Permission Handshake Mechanism}.We introduced a dual-factor decision model $\Phi(S, R)$, where $S$ is user strictness (0, 5, 10) and $R$ is request sensitivity (0-10). The system determines actions (Allow, Negotiate, Block) based on the risk score ($S \times R$). We simulated 360 interactions with three distinct agents:
\begin{itemize}
    \item \textbf{FinTech Agent ($R=9$):} Legitimate high-value service (requires asset data).
    \item \textbf{DataBroker Agent ($R=8$):} Malicious profiler (attempts to scrape identity).
    \item \textbf{Academic Agent ($R=2$):} Legitimate low-risk service (public interests).
\end{itemize}

\begin{table}[H]
\centering
\caption{Impact of Strictness Parameter (S) on Privacy and Utility}
\label{tab:privacy_robustness}
\begin{tabular}{@{}lllll@{}}
\toprule
\textbf{Metric} & \textbf{Baseline} & \textbf{Ours (S=0)} & \textbf{Ours (S=5)} & \textbf{Ours (S=10)} \\ \midrule
High-Risk Protection Rate ($P_{safe}$) & 0.00\% & 97.50\% & 97.50\% & \textbf{100.00\%} \\
Service Availability ($U_{service}$) & 100.00\% & 100.00\% & \textbf{100.00\%} & 50.00\% \\ Token Cost ($C_{token}$) & 0 & 1272.20 & 1233.22 & 409.17 \\ \bottomrule
\end{tabular}
\end{table}

The experimental results reveal a critical trade-off between privacy security and service utility. Without the Digital Avatar, the system exhibited a complete failure in protection (0\%), exposing all private data to malicious profiling. In contrast, the Avatar architecture introduces a safety baseline; even at the most permissive setting ($S=0$), the Hard Rule ($R \ge 8$) successfully forced high-risk requests into negotiation, preventing catastrophic leakage (97.5\% protection), albeit with higher token costs due to desensitization overhead. Crucially, the configuration at $S=5$ represents a Golden Balance, maintaining 100\% Service Availability for legitimate high-value agents (FinTech) while safeguarding 97.5\% of privacy. However, excessive strictness ($S=10$) leads to over-protection, where perfect security (100\%) comes at the cost of a significant drop in availability (50\%) due to false positives, highlighting the necessity of adaptive governance over rigid blocking.

\section{Conclusion}
\label{sec:5}
This paper proposes the \textbf{SoDA}, a concept that fundamentally revolutionizes the existing interaction paradigm within the Agentic Web. By addressing the critical challenges of data lock-in and cognitive overload, our architecture establishes a new standard for user sovereignty. Our rigorous empirical evaluation confirms that the orthogonal decoupling of storage and computation transforms user memory from a platform-bound resource into a portable efficiency multiplier. Specifically, SMP eliminates the need for exploratory interaction loops, doubling task execution speeds, while the \textit{Dual-Factor Adaptive Routing} mechanism enables a Golden Balance between strict privacy protection and service availability. By reducing the interaction friction coefficient to a negligible 0.05, this paradigm effectively liberates users from the cognitive burden of tool management and prompt engineering. Consequently, the SoDA proves to be not merely a privacy shield, but the foundational infrastructure required to realize an efficient, secure, and truly decentralized Agentic Web.

Future work will focus on conducting longitudinal human-subject studies to validate the long-term interaction experience beyond simulations. Furthermore, we aim to extend the proposed UPDL into a standardized universal protocol and explore decentralized reputation mechanisms. This will empower the SoDA to evolve from a static privacy shield into an active ecosystem orchestrator, securely navigating the complex collaboration and trust challenges of the open Agentic Web.

\bibliography{main}
\bibliographystyle{rlc}

\end{document}